\documentclass[conference]{IEEEtran}
\IEEEoverridecommandlockouts
\usepackage{cite}
\usepackage{amsmath,amssymb,amsfonts}
\usepackage{algorithmic}
\usepackage{graphicx}
\usepackage{textcomp}
\usepackage{xcolor}
\def\BibTeX{{\rm B\kern-.05em{\sc i\kern-.025em b}\kern-.08em
    T\kern-.1667em\lower.7ex\hbox{E}\kern-.125emX}}

\usepackage{pgfpages}
\pgfpagesuselayout{resize to}[a4paper]   
    
\begin{document}

\title{An Intelligent Scheme for Uncertainty Management of Data Synopses Management in Pervasive Computing Applications%\\
}

\author{\IEEEauthorblockN{Kostas Kolomvatsos}
\IEEEauthorblockA{\textit{Department of Informatics and Telecommunications} \\
\textit{University of Thessaly}\\
Papasiopoulou 2-4, 35131 Lamia Greece \\
kostasks@uth.gr}
%\and
%\and
%\and
%\and

}

\maketitle

\begin{abstract}
Pervasive computing applications deal with the incorporation of 
intelligent components around end users to facilitate their activities.
Such applications can be provided upon the vast infrastructures of 
Internet of Things (IoT) and Edge Computing (EC).
IoT devices collect ambient data transferring them towards the EC and Cloud for further processing.
EC nodes could become the hosts of distributed datasets 
where various processing activities take place.
The future of EC involves numerous nodes interacting 
with the IoT devices and themselves in a cooperative manner to realize the desired processing.
A critical issue for concluding this cooperative approach 
is the exchange of data synopses to have EC nodes informed about the data 
present in their peers.
Such knowledge will be useful for decision making related to the execution of processing activities.
In this paper, we propose n uncertainty driven model for the exchange of data synopses.
We argue that EC nodes should delay the exchange of synopses especially when no significant differences with historical values are present.
Our mechanism adopts a Fuzzy Logic (FL) system to decide when there is 
a significant difference with 
the previous reported synopses to decide the exchange of the new one.
Our scheme is capable of alleviating the network from numerous messages 
retrieved even for low fluctuations in synopses.
We analytically describe our model and evaluate it through a large set of experiments.
Our experimental evaluation targets to detect the efficiency of the approach based on the elimination of 
unnecessary messages while keeping immediately informed peer nodes for significant 
statistical changes in the distributed datasets. 
\end{abstract}


\section{Introduction}
The Internet of Things (IoT) provides a huge infrastructure where
numerous devices can interact with end users and their environment to 
collect data and perform simple processing activities 
\cite{najam}.
IoT devices can report their data to the Edge Computing 
(EC) infrastructure and Cloud for further processing.
As we move upwards from the IoT to the EC and Cloud, 
we meet increased computational resources, however, accompanied by increased latency. 
EC has been proposed as the paradigm adopted to be close to the IoT platform and end users involving 
increased processing capabilities (compared to IoT) to reduce the latency we enjoy when relying to
Cloud. 
EC deals with the provision of storage and processing capabilities 
from various heterogeneous devices \cite{najam}.
EC nodes can become the hosts of distributed datasets formulated by the reports of IoT devices.
There, we can incorporate advanced services to produce knowledge and analytics to immediately 
respond to any request, thus, supporting real time applications.

The aforementioned distributed datasets become the 
subject of numerous requests having the form of processing tasks or queries.
Various research efforts 
study the selection of data hosts based 
on their available memory and battery levels \cite{amrutha} to perform the 
execution of tasks/queries.
The future of EC involves nodes that are capable of cooperating to 
perform the desired tasks/queries. 
Under this `cooperative' perspective,
having a view on the statistics of the available datasets may assist 
in the definition 
of efficient tasks/query allocations.
For instance, an EC node may decide to offload a task/query for various performance reasons.
The research community has already proposed 
data migration \cite{bellavista} as a solution to efficiently respond to requests.
However, migrating huge volumes of data may jeopardize the stability of the network
due to the increased bandwidth required to perform such an action.
A solution is the offloading of tasks/queries, however,
the allocation decision should be based on the 
data present in every peer node.
The decision making should be realized upon the statistics of the available datasets
to conclude the most appropriate allocation.

In this paper, we focus on the autonomous nature of EC nodes and 
propose a scheme for distributing data synopses to peers.
We argue for the dissemination of the synopsis of each dataset
to have the insight on the data present in peers.
We propose the monitoring of synopses updates and detect when 
a significant deviation (i.e., the magnitude) with the previous reported 
synopsis is present.
We define an uncertainty driven model under the principles 
of Fuzzy Logic (FL) \cite{fuzzy} to decide when an EC node should 
distribute the synopsis of its dataset.
The uncertainty is related to the `threshold' (upon the differences 
of the available data after getting reports from IoT devices) over which 
the node should disseminate the current synopsis.
We monitor the `statistical significance'
of synopses updates before we decide to distributed them in the network.
This way, we want to avoid the continuous distribution of synopses especially
when no significant difference is present.
We consider the trade off between the frequency of the distribution and the 
`magnitude' of updates. We can accept the limited freshness of updates for gaining
benefits in the performance of the network.
Our FL-based decision making mechanisms adopts 
Type-2 FL sets to cover not only the uncertainty in the decision making but also in the definition 
of membership functions for every FL set.
We apply our scheme upon past, historical observations (i.e., synopses updates)
as well as upon future estimations.
We adopt a forecasting methodology for estimating the `trend' in synopses updates. 
Both, the view on the past and the view on the future are fed into our Type-2 FL System (T2FLS)
to retrieve the \textit{Potential of Distribution} (PoD).
Two PoD values (upon historical values and future estimations) are smoothly aggregated through a 
geometrical mean function \cite{mesiar} to finally decide the dissemination action.
Our contributions are summarized by the following list:
\begin{itemize}
	\item We provide a monitoring mechanism for detecting the magnitude of the updated synopses;
	\item We deliver a forecasting scheme for estimating the future realizations of 
	data synopses; 
	\item We describe and analyze an uncertainty driven model for detecting the appropriate time to distribute data synopses to peer nodes;
	\item We report on the experimental evaluation of the proposed models through a large set of simulations.
\end{itemize}

%organization
The paper is organized as follows.
Section \ref{related} presents the related work while Section 
\ref{preliminaries} formulates our problem and provides the 
main notations adopted in our model.
In Section \ref{uncertainty}, we present the envisioned mechanism and 
explain its functionalities.  
In Section \ref{evaluation}, we deliver our experimental evaluation
and conclude the paper in Section \ref{conclusions} by presenting 
our future research directions.

\section{Related Work}
\label{related}
%resource management in edge nodes
A significant research subject in EC is resource management.
It is critical to adopt efficient techniques
to resources allocation either in the form of
scheduling or in the form of the allocation of tasks/queries to the appropriate resources.
The ultimate goal is to increase the performance and 
facilitate the desired processing and timely provide responses. 
Currently, EC nodes adopt the following models to 
execute tasks/queries \cite{wang}:
(i) through an aggregation model where data coming from multiple devices can be collected and pre-processed in an edge node \cite{ref17}.
Data are locally processed before they are transferred to the Cloud limiting the time for the provision of the final response; 
(ii) through a `cooperative' model where EC nodes can interact with IoT devices having processing capabilities to
offload a subset of tasks \cite{ref21}. 
Devices and EC nodes should exhibit a low distance, otherwise,
their interaction may be problematic;
(iii) through a `centralized' approach where edge nodes act as execution points for tasks/queries offloaded by IoT devices \cite{ref26}. 
EC nodes exhibit higher computational capabilities than IoT devices 
and can undertake the responsibility of performing `intensive' tasks, however, under the danger of being overloaded. 
Current research efforts related to tasks/queries management at EC nodes 
focus on caching \cite{ref32}, 
context-aware web browsing \cite{ref33} 
and video pre-processing \cite{ref34}.
A number of efforts try to 
deal with the resource management problem
\cite{anglano}, \cite{Cherrueau}, \cite{Shekhar}, \cite{wang}.
Their aim is to address the challenges on how we can 
offload various tasks/queries and data to EC nodes taking into consideration 
a set of constraints, e.g., time requirements, communication 
needs, nodes' performance, the quality of the provided responses and so on
and so forth. 

The current form of the IoT and EC involves 
numerous devices that collect, report and process data.
Due to the huge amount of data, data synopses can be useful into a variety 
of IoT/EC applications. 
Synopses depict the `high' level description 
of data and represent their statistics \cite{aggarwal}.
The term `synopsis' usually refers in 
(i) approximate query estimation \cite{Chakrabarti}: 
we try to estimate the responses given the query. This should be performed
in real time. The processing aims to estimate the data that will 
better `match' to the incoming queries;
(ii) approximate join estimation \cite{alon}, \cite{dobra}: we try to estimate 
the size of a join operation that is significant in `complex' operations over the 
available data;
(iii) aggregates calculation \cite{Charikar}, \cite{Cormode}, \cite{Gehrke}, \cite{Manku}: the aim is to provide aggregate statistics over the available data;
(iv) data mining mechanisms \cite{Aggarwal1}, \cite{Aggarwal2}, \cite{Schweller}: some services may demand for synopses instead of the individual data points, e.g., clustering, classification.
In any case, the adoption of data synopses aims at the processing of 
only a subset of the actual data.
Synopses act as `representatives' of data and usually involve 
summarizations or the selection of a specific subset
\cite{Lakshmi}.
These limited representations reduce the need 
for increased bandwidth of the network and 
can be easily transferred in the minimum possible time.
Some synopses definition techniques involve
sampling \cite{Lakshmi},
load shedding 
\cite{Babcock1}, \cite{Tatbul},
sketching \cite{Babcock}, \cite{Muthukrishnan} and 
micro cluster based summarization \cite{Aggarwal1}.
Sampling is the easiest one targeting to the probabilistic selection of a subset of the actual data.
Load shedding aims to drop some data when the system identifies a high load, thus, to 
avoid bottlenecks.
Sketching involves the random projection of a subset of features that describe the data 
incorporating mechanisms for the 
vertical sampling of the stream.
Micro clustering targets to the management of the multi-dimensional aspect of any data stream 
towards to the processing of the 
data evolution over time.
Other statistical techniques 
are histograms and wavelets \cite{aggarwal}.

\section{Preliminaries and Problem Description}
\label{preliminaries}
We consider a set of $N$ EC nodes 
$\mathcal{N} = \left\lbrace n_{1}, n_{2}, \ldots, n_{N} \right\rbrace$, with their corresponding datasets 
$\mathcal{D} = \left\lbrace D_{1}, D_{2}, \ldots, D_{N} \right\rbrace$.
Every dataset $D_{i} = \{\mathbf{x}_{j}\}_{j=1}^{m_{j}}$ contains $m_{j}$ real-valued contextual multidimensional data vectors 
$\mathbf{x} = [x_{1}, x_{2}, \ldots, x_{d}]^{\top} \in \mathbb{R}^{d}$
of $d$ dimensions. 
Every dimension refers to a contextual attribute (e.g., temperature, humidity).
Contextual data vectors are reported by IoT devices that capture them through interaction with their environment.
Contextual data vectors become the basis 
for knowledge extraction for every $n_{i}$. 
An arbitrary methodology is adopted like a regression analysis, classification tasks, the estimation of multivariate and/or uni-variate histograms per attribute, non-linear statistical dependencies between input attributes and an application-defined output attribute, clustering of the contextual vectors,
etc. 
Without loss of generality, we assume the online knowledge extraction model as the \textit{statistical synopsis} $\mathcal{S}$. $\mathcal{S}$ is represented by $l$-dimensional vectors, i.e., 
$\mathbf{s} = [s_{1}, s_{2}, \ldots, s_{l}]^{\top} \subset \mathbb{R}^{l}$.
A statistical synopsis $\mathcal{S}_{i}$ is the summarization of 
$D_{i}$ located at $n_{i}$. 
We obtain $N$ data synopses $\mathcal{S}_{1}, \ldots, \mathcal{S}_{N}$ represented via their synopsis vectors. 
Given synopses, EC nodes, initially, are responsible for 
maintaining their up-to-date synopses as the underlying data may change (e.g., concept drift). 
Additionally, EC nodes try to act in a cooperative manner and decide to 
exchange their synopses regularly. 
Through this approach, EC nodes 
can have a view on the statistical properties of datasets present in their peers.
Based on that, we can gain advantage on preforming decision making for allocating tasks/queries and deliver analytics taking into consideration the data synopses distributed at the EC network.

Arguably, there is a trade off between the 
communication overhead and the `freshness' of synopses delivered to peer nodes.
EC nodes can share up-to-date synopses every time a change in the underlying data is realized
at the expense of flooding the network with numerous messages. Recall, that EC nodes
are connected with IoT devices that are continuously reporting 
data vectors in high rates.
However, in this case, peer nodes enjoy fresh information
increasing the performance of decision making.
The other scenario is to postpone the delivery of synopses,
i.e., to reduce the sharing rate 
expecting less network overhead in light of `obsolete' synopses.
In this paper, we go for the second scenario and try to detect the appropriate time 
to deliver a synopsis to peer nodes. 
The target is to optimally limit the messaging overhead.
The idea is to let EC nodes to decide the `magnitude' of the 
collected statistical synopsis before they decide a dissemination action.
Obviously, there is uncertainty around the amount of magnitude that should be realized 
before we fire a dissemination action.
We propose an uncertainty driven mechanism, i.e., out T2FLS that results the PoD upon 
past synopsis observations and its estimated values. 
In any case, EC nodes are forced to disseminate synopses 
at pre-defined intervals even if no delivery decision is the outcome from 
our model.
We have to notice that, to avoid bottlenecks in the network, we consider the pre-defined intervals to differ among the group of EC nodes.
This `simulates' a load balancing approach avoiding to have too many EC nodes disseminating their synopses at the same time.

Our T2FLS is fed by the most recent $\mathcal{S}$ as well as 
with its future realizations. 
Every EC node monitors significant changes in the local synopsis 
as more contextual data are received from IoT devices.
Based on this local monitoring, implicitly, we incorporate 
into the network edge the necessary `randomness' in the 
conclusion of the final decision, thus, potentially avoiding network flooding.
The discussed `randomness' is enhanced by different 
data arriving to the 
available nodes and their autonomous decision making. 
Such `randomness' can assist in limiting the possibility of deciding the delivery of synopses at the same time, thus, we can limit the possibility of overloading the network. 
% a few words on the updates of the data synopses
Let us consider that at the discrete time instance $t$ a new data vector 
arrives in $n_{i}$.
Afterwards, the corresponding synopsis
$\mathbf{s}_{i}$ should be updated to conclude the 
new $\mathbf{s}^{t}_{i}$.
Let $\mathbf{e}_{t}$ be the difference over 
the current, last sent synopsis $\mathbf{s}_{i}$ and the new, the updated one,
$\mathbf{s}^{t}_{i}$.
We call this error/difference
as the \textit{update quantum}, i.e., the magnitude of the difference between $\mathbf{s}_{i}$ and $\mathbf{s}^{t}_{i}$.
$n_{i}$ calculates $\mathbf{e}_{t}$ at consecutive 
time steps.
$\mathbf{e}_{t}$, in a simplistic way, can be concluded by adopting the 
\textit{sum of differences} between two consecutive synopsis
for every dimension.
Obviously, we can adopt any desired synopses realization technique as mentioned above.
$\mathbf{e}_{t}$ may be positive or negative, i.e.,  
a new vector can increase or decrease the value of each dimension.
For facilitating our calculations, we are based on the
absolute value for any difference.
EC nodes should delay the delivery of $\mathbf{s}^{t}_{i}$
until they see that a significant difference, i.e., 
a high \textit{magnitude} of $\mathbf{e}_{t}$
is present.
In that time, it is necessary to have the peer nodes informed 
about the new status of the local dataset.
We define the \textit{update epoch} as the time between disseminating two consecutive 
synopsis updates. 
The update epoch is realized at pre-defined intervals,
$T, 2T, 3T, \ldots$ ($T >0$).
To describe our solution, we focus on an individual interval, 
e.g., $[1, 2, \ldots, T]$.
At each $t \in [1, 2, \ldots, T]$, EC nodes 
check the last $\mathbf{e}$ realizations and feed them into our T2FLS to 
see if they excuse the initiation of the dissemination process.
This action should be realize till $T$.
If no dissemination decision is made till $T$,
EC nodes start the dissemination no matter the observed magnitude.
EC nodes also `reason' over the time series of update quanta 
$\left\lbrace \mathbf{e}_{t} \right\rbrace$ with $t=1, 2, \ldots, T$.
EC nodes `project' the time series to the future through the adoption of 
a forecasting technique. Again, the projection of update quanta is fed into the T2FLS to generate 
the PoD upon the future estimations of $\mathbf{e}$.
The final goal is to accumulate as much as possible $\mathbf{e}$
before we decide the dissemination action.
When the accumulated magnitude is relatively high,
EC nodes decide to `stop' the monitoring process, disseminate the updated
synopsis and `start off' a new monitoring/update epoch.

\section{Uncertainty Driven Proactive Synopses Dissemination}
\label{uncertainty}
\textbf{The Proposed Fuzzy Reasoning Process}. For describing the proposed T2FLS, we borrow the notation of our previous efforts (in other domains)
presented in \cite{kolomvatsos2017}, \cite{kolomvatsos2016}.
T2FLS is adopted locally at every node
at $t$ 
by fusing 
the past $\mathbf{e}_{t}$ observations 
and future $\mathbf{e}_{t}$ realizations. $\mathbf{e}_{t}$
is adopted as the indication
whether the current update quanta 
significantly 
deviate from 
their past and future short-term trends. 
The envisioned fusion of update quanta is 
achieved through a finite set of
\textit{Fuzzy Inference Rules} (FIRs).
FIRs incorporate and `combine' past quanta or future estimations 
(two different processes) to reflect 
the $PoD$. Actually, we `fire' two consecutive times the T2FLS for 
the last three (3) quanta realizations, i.e., $\mathbf{e}_{t-2}, \mathbf{e}_{t-1}, \mathbf{e}_{t}$ and the future three (3) 
quanta estimations, i.e., $\mathbf{e}_{t+1}, \mathbf{e}_{t+2}, \mathbf{e}_{t+3}$.
Our T2FLS, 
defines 
the fuzzy 
knowledge base 
for every $n_{i}$, 
e.g., a set of FIRs like: 
`\textit{when the past/future quanta exhibit a significant 
difference from the last synopsis delivery,  
the $PoD$ for initiating the delivery of the new synopsis might 
be also high}'. 
We rely on Type-2 FL sets as 
the `typical' Type-1 sets 
and the FIRs defined upon them involve uncertainty due to
partial knowledge in representing 
the output of the inference \cite{ref45}.
The limitation 
in a Type-1 FL system is 
on handling uncertainty 
in representing knowledge 
through FIRs \cite{ref28}, \cite{ref45}.
In such cases, uncertainty 
is observed 
not only in the environment, 
e.g., we classify 
the $PoD$ as `high', 
but also on the 
description of the term, e.g., `high', 
itself. 
In a T2FLS,  
membership functions  
are themselves `fuzzy',
which leads 
to the definition of FIRs 
incorporating such uncertainty \cite{ref45}. 

FIRs refer to a 
non-linear mapping
between three inputs:
(i) when focusing on the past quanta, we take into consideration the following
as the inputs into the T2FLS: 
$\mathbf{e}_{t-2}, \mathbf{e}_{t-1}, \mathbf{e}_{t}$;
(ii) when focusing on the future quanta, we take into consideration the following
as the inputs into the T2FLS: 
$\mathbf{e}_{t+1}, \mathbf{e}_{t+2}, \mathbf{e}_{t+3}$.
The outputs are $PoD_{p}$ \& $PoD_{f}$, respectively.
The antecedent part of FIRs 
is a (fuzzy) conjunction of inputs and 
the consequent part of the FIRs 
is the $PoD$ indicating 
the belief that an event \textit{actually} occurs.
The proposed FIRs have 
the following structure:
\noindent
\textbf{IF} $\mathbf{e}_{t-2}$ is $A_{1k}$ \textbf{AND} $e_{\mathbf{e}_{t-1}}$ is $A_{2k}$ \textbf{AND} $\mathbf{e}_{t}$ is $A_{3k}$ \\
\noindent 
\textbf{THEN} $PoD_{p}$ is $B_{k}$,
\noindent

\noindent
\textbf{IF} $\mathbf{e}_{t+1}$ is $A_{1k}$ \textbf{AND} $e_{\mathbf{e}_{t+2}}$ is $A_{2k}$ \textbf{AND} $\mathbf{e}_{t+3}$ is $A_{3k}$ \\
\noindent 
\textbf{THEN} $PoD_{f}$ is $B_{k}$,
\noindent

where $A_{1k}, A_{2k}, A_{3k}$ and $B_{k}$ are 
membership functions for the $k$-th FIR 
mapping $\mathbf{e}_{i}, \mathbf{e}_{j}, \mathbf{e}_{k}$ and $PoD_{m}$,
$i \in \left\lbrace t-2, t+1 \right\rbrace$,
$j \in \left\lbrace t-1, t+2 \right\rbrace$,
$k \in \left\lbrace t, t+3 \right\rbrace$ and
$m \in \left\lbrace p, f \right\rbrace$.
For FL sets, we characterize their 
values through the terms: \textit{low}, \textit{medium}, and \textit{high}. 
The structure of  
FIRs in the proposed T2FLS 
involve linguistic terms, e.g., \textit{high},  
represented by two membership functions, i.e.,
the 
\textit{lower} and the \textit{upper} bounds \cite{ref44}. 
For instance, the term `\textit{high}' 
whose membership for $x$ is a number $g(x)$, 
is represented by two membership functions defining the 
interval $[g_{L}(x), g_{U}(x)]$.
This interval 
corresponds to a lower and an upper membership function 
$g_{L}$ and $g_{U}$, respectively 
(e.g., the membership of $x=0.25$ can be in the interval $[0.05, 0.2]$). 
The interval areas $[g_{L}(x_{j}), g_{U}(x_{j})]$ for each  $x_{j}$ 
reflect the uncertainty in defining the term, e.g., `\textit{high}',  
useful to determine the exact membership function 
for each term. 
Obviously, if $g_{L}(x) = g_{U}(x), \forall x$, 
we obtain a FIR in a Type-1 FL system. 
The interested reader could refer to \cite{ref44}
for information on reasoning under 
Type-2 FIRs.

\textbf{Time Series Forecasting}. Exponential smoothing \cite{holt} is a time series 
estimation methodology for univariate data. The method can be easily extended 
to detect the trend or seasonal components on data. We have to notice that we adopt the specific methodology as it is fast, is easily adapted to frequent changes of data and 
performs better than other techniques (e.g., moving average), especially for a short-term time horizon.
Exponential smoothing 
is similar to the weighted sum of past observations, however,
it adopts a decay factor for decreasing weights based on the index of the 
past observation.
In our model, we adopt the double exponential smoothing for estimating 
$\mathbf{e}_{t+1}, \mathbf{e}_{t+2}, \mathbf{e}_{t+3}$ based on 
all the available/calculated synopsis quanta.
Recall that $\mathbf{e}_{t}$ exhibits the statistical difference (e.g., the sum)
of the current and the previous disseminated synopsis.
Hence, we can rely on the univariate scenario hiding all the statistics 
for each individual dimension. This is an `abstraction' strategically adopted in our
model.
In the first place of our future research plans is the 
application of forecasting techniques for each individual 
dimension, then, aggregating them to derive the final estimated update
quanta. 
Let all the available quanta be $\mathbf{e}_{1}, \mathbf{e}_{2}, \ldots, \mathbf{e}_{t}$.
When applying the double exponential smoothing model, the following equations hold true:
$v_{j} = \alpha \mathbf{e}_{j} + (1 - \alpha) (v_{j-1} + b_{j-1})$, 
$b_{j} = \beta (v_{j} - v_{j-1}) + (1-\beta) b_{j-1}$, 
where
$v_{1} = \mathbf{e}_{1}$ \& $b_{1} = \mathbf{e}_{2} - \mathbf{e}_{1}$.
Additionally, $\alpha \in (0,1)$ is the data smoothing factor and $\gamma \in (0,1)$ is the trend smoothing factor.
The method adopts two smoothing 
factors to control the decay of data and the decay of the influence of the change in the trend of data.
For performing a forecasting for additional data in the future, we adopt the following equation:
$v_{j+k} = v_{j} + k b_{j}$.
Based on the above, we can easily get 
$\mathbf{e}_{t+1}, \mathbf{e}_{t+2}, \mathbf{e}_{t+3}$
quanta fed into our T2FLS to retrieve
the $PoD_{f}$.

\textbf{The Decision Making Mechanism}. 
Our T2FLS is responsible to deliver $PoD_{p}$ and $PoD_{f}$ based on the most recent past observations for $\mathbf{e}$ and the estimated future realizations. Hence, we have to combine the experience 
of an EC node for the update quanta as already recorded with its
insight on the future. 
We propose a simple aggregation process for $PoD_{p}$ and $PoD_{f}$
(to be realized in real time) based on the geometric mean
\cite{mesiar}.
The following equation holds true:
$G(PoD_{p}, PoD_{f}) = \left( \prod_{i=1}^{2} PoD_{i} \right)^{1/2}$,
with $i \in \left\lbrace p, f \right\rbrace$.
We rely on the geometric mean instead of other methodologies 
as it deals with all the inputs (i.e., our PoD values)
and is not affected by extreme low or high values.
Additionally, we want to incorporate into our decision making 
a `strict' approach, i.e., when a PoD value is zero then the final outcome is zero as well. Through this approach we try to 
be sure about the magnitude of update quanta 
before we decide to initiate the dissemination action.
Finally, when $G>theta$, we initiate the dissemination action.
$\theta$ is a pre-defined threshold that `dictates' when 
an EC node should pursue the exchange of synopsis.

\section{Experimental Setup and Evaluation}
\label{evaluation}
\textbf{Setup and Performance Metrics}.
We report on the performance of our \textit{Uncertainty Driven
Dissemination Model} (UDDM) and compare it 
with other baseline models and schemes proposed in the relevant literature.
Initially, we focus on the  
percentage of $T$ that our model spends till the final decision.
The $\phi$ metric is defined as follows:
$\phi = \frac{1}{E} \sum \left\lbrace \frac{t^{*}}{T} \right\rbrace_{i=1}^{E}$
where $t^{*}$ is the time when the dissemination actions is decided,
$E$ is the number of  
experiments and 
$i$ depicts the index of every experiment. 
When $\phi \to 1$ means that the 
proposed model spends the entire interval $T$ 
to conclude a final decision.
When $\phi \to 0$, our model manages to 
conclude immediately the dissemination action.
Additionally, we define 
the metric $\delta$ i.e., 
$\delta = \frac{1}{E} \sum \left\lbrace |\mathbf{s}^{t^{*}} - \mathbf{s}| \right\rbrace_{i=1}^{E}$.
$\delta$ represents the average magnitude 
of the difference between the current and the 
new synopses.
Through the use of $\delta$, we want to present the 
ability of the proposed model to `react' even in 
limited changes in the updated synopses 
(we target a $\delta \to 0$).
The magnitude is calculated 
at $t^{*}$.
The ability of the proposed system to avoid the overloading of the network and limiting the required number of messages is exposed by  
$\psi$. 
The following equation holds true:
$\psi = \frac{T}{|t^{*}|_{t^{*} \in [1,T]}}$ ($\psi \in [0,T]$)
where $|t^{*}|_{t^{*} \in [1,T]}$ represents the number of 
times that the model stops 
in the interval $[1,T]$.
When $\psi \to 1$ means that the proposed model 
stops frequently, thus, multiple messages conveying the
calculated synopses are transferred through the network.
When $\psi \to T$ means that our model does not stop frequently, thus, the calculated synopses are delivered after the expiration of the window $T$.
For our experimentation, we adopt the dataset 
presented in Intel Berkeley Research Lab \cite{chu}. 
It contains measurements from 54 sensors deployed 
in a lab. We get the available measurements and simulate the provision 
of context vectors to calculate the synopses and the update quanta (they are relized
in the interval $[0, \infty]$)
in a sequential order. 
We also pursue a comparative assessment for the UDDM with:
\textbf{(i)} a baseline model (BM) that 
disseminates synopses when any change is observed over 
the incoming data; 
\textbf{(ii)} the Prediction based Model (PM)
\cite{martin}: 
PM proceeds with the stopping decision only when 
the estimation of the future update quanta violates 
a threshold.
We perform simulations for $E = 1,000$ and $T \in \left\lbrace 10, 100 1000 \right\rbrace$. 
In every experiment, we run the UDDM and get numerical results related to the mean values of the aforementioned metrics (we adopt $\theta \in \left\lbrace 0.60, 0.75 \right\rbrace$ for the UDDM and the PM). 

\begin{figure}[h]
\centerline{\includegraphics[width=85mm,scale=0.95]{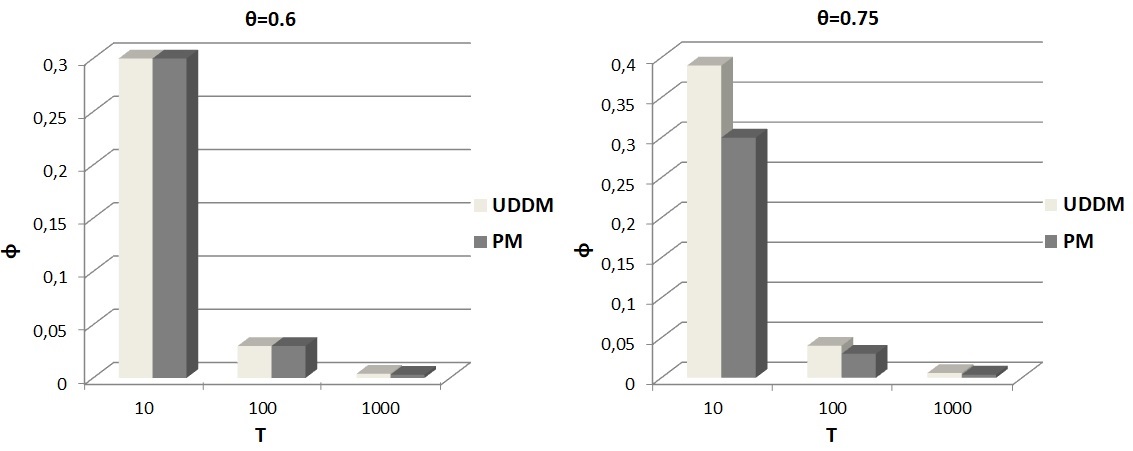}}
\caption{Comparative results for the $\phi$ metric.}
\label{fig1}
\end{figure}

\textbf{Performance Assessment}. In Fig. \ref{fig1}, we present our results for the $\phi$ metric. We observe that the adoption of a low $\theta$ (threshold for deciding the 
dissemination action) and a low $T$
(deadline to conclude the distribution of synopses) lead to an increased time for the final decision. Even in that case, the required time is around 30\% of the total deadline $T$. 
When $\theta$ and $T$ are high, the percentage of $T$ devoted to conclude the dissemination decision is very low.
Actually, the proposed system manages to deal with the final decision as soon as it detects that update quanta are aggregated over time even in small amounts. This can be realized in early monitoring rounds due to the dynamic nature of the incoming data.
Recall that we adopt a time series that consists of sensory data retrieved by a high number of devices that are, generally, characterized by their dynamic nature.

In Table \ref{tab:table1}, we present our results related to the $\delta$ metric,
i.e., the update quanta at the time when the dissemination action is decided.
We observe that the UDDM requires a higher magnitude than the BM and the PM before it concludes the dissemination action. This stands for both experimental scenarios, i.e.,
$\theta \in \left\lbrace 0.6, 0.75 \right\rbrace$.
In general, there is an increment in $\delta$ as $T$ increases. Additionally, the PM exhibits the lowest $\delta$ outcome, i.e., 
the update quanta for which a dissemination action is decided. These result present the `attitude' of the proposed model to wait and aggregate update quanta in order to alleviate the network from an increased number of messages.    

\begin{table}[h]
\caption{Experimental outcomes for the $\delta$ metric} 
\label{tab:table1}
\begin{tabular}{l|lll|lll}
\hline \hline
     & \multicolumn{3}{c|}{\textbf{$\theta = 0.6$}}                                                           & \multicolumn{3}{c}{\textbf{$\theta = 0.75$}}                                                          \\
\multicolumn{1}{l|}{\textbf{T}} & \multicolumn{1}{c}{\textbf{UDDM}} & \multicolumn{1}{c}{\textbf{BM}} & \multicolumn{1}{c|}{\textbf{PM}} & \multicolumn{1}{c}{\textbf{UDDM}} & \multicolumn{1}{c}{\textbf{BM}} & \multicolumn{1}{c}{\textbf{PM}} \\
\hline \hline
10   & 16.42    & 13.87  & 13.87   & 15.82    & 12.05   & 8.61  \\
100   & 20.95 & 17.76   & 17.02   & 17.60    & 15.59    & 13.92  \\
1,000   & 19.62  & 17.68   & 16.34 & 20.55  & 17.79  & 16.63 \\
\hline \hline                    
\end{tabular}
\end{table}

Table \ref{tab:table2} depicts our experimental evaluation related to the 
$\psi$ metric. 
We observe that the UDDM demands for less dissemination messages compared with the BM \& PM.
As $T \to 1,000$, BM and PM exhibit an increased number messages. Approximately, they deliver the update quanta at every monitoring round.
The proposed model decides the dissemination of messages 
every 2.5 (approximately) monitoring rounds 
for the experimental scenario where $\theta = 0.6$.
When $\theta = 0.75$, we observe an increment in the dissemination activity, i.e., for every 1.5 monitoring rounds.

\begin{table}[h]
\caption{Experimental outcomes for the $\psi$ metric} 
\label{tab:table2}
\begin{tabular}{l|lll|lll}
\hline \hline
     & \multicolumn{3}{c|}{\textbf{$\theta = 0.6$}}                                                           & \multicolumn{3}{c}{\textbf{$\theta = 0.75$}}                                                          \\
\multicolumn{1}{l|}{\textbf{T}} & \multicolumn{1}{c}{\textbf{UDDM}} & \multicolumn{1}{c}{\textbf{BM}} & \multicolumn{1}{c|}{\textbf{PM}} & \multicolumn{1}{c}{\textbf{UDDM}} & \multicolumn{1}{c}{\textbf{BM}} & \multicolumn{1}{c}{\textbf{PM}} \\
\hline \hline
10   & 2.5    & 1.42  & 1.66   & 1.66    & 1.66   & 1.42  \\
100   & 2.5 & 1.2   & 1.13   & 1.47    & 1.29    & 1.23  \\
1,000   & 2.09  & 1.2   & 1.12 & 1.46  & 1.24  & 1.15 \\
\hline \hline                    
\end{tabular}
\end{table}

\section{Conclusions}
\label{conclusions}
One of the most significant research subjects at the Edge Computing (EC) is the management of data coming from devices active in the Internet of Things (IoT). The reason is that the IoT will become the core infrastructure for hosting future applications. 
This means that IoT should be supported by efficient services that build upon the collected data.
Data management at the EC provides many benefits with the most important of them to be the minimization of the latency in the provision of responses in tasks/queries.
A set of EC nodes can be adopted to become the hosts of the collected data and 
the processing entities being very close to end users and the IoT infrastructure.
We argue on the cooperative approach that EC nodes should follow in order to conclude advanced processing mechanisms.
Under this cooperative model, EC nodes could exchange 
the statistical synopses of their datasets to support the efficient decision making of their peers towards realizing the most productive tasks/queries allocations. This decision making mechanism is `fired' when an EC node decides to offload the incoming tasks/queries.
In this paper, we present a novel, uncertainty driven 
model to reason over the appropriate time to exchange data synopses. 
Our aim is to provide a scheme that minimizes the number of messages circulated in the network, however, without jeopardizing 
the freshness of the exchanged statistical information.
We discuss our model adopting the principles of Fuzzy Logic (FL)
and present the relevant formulations.
EC nodes monitor their data and decide when it is
the right time to deliver the current data synopsis. 
Our experimental evaluation shows that the proposed scheme can efficiently assist 
in the envisioned goals being evidenced by 
numerical results. 
In the first place of our future research plans, it is to incorporate a deep machine learning model in the uncertainty management scheme. 
Additionally, we want to involve more parameters in the decision making 
mechanism like a `snapshot' of the current status of every EC node.

%\section{References}


\begin{thebibliography}{00}
\bibitem{amrutha}
Amrutha, S. et al., 'Data Dissemination Framework for IoT based Applications', Indian Journal of Science and Technology, 9(48), 2016, pp. 1--5.

\bibitem{Aggarwal1}
Aggarwal, C., Han, J., Wang J., Yu, P., 'A Framework for Clustering Evolving Data Streams', VLDB Conference, 2003.

\bibitem{Aggarwal2}
Aggarwal, C., Han, J., Wang, J., Yu, P., 'On-Demand Classification of Data Streams', ACM KDD Conference, 2004. 

\bibitem{aggarwal}
Aggarwal, C., Yu, P., 'A Survey of Synopsis Construction in Data Streams', ch. in 'Data Streams, Models and Algorithms', ed. Aggarwal, C., Springer Science \& Business Media, 2007.

\bibitem{alon}
Alon, N., Gibbons P., Matias Y., Szegedy, M.,  'Tracking Joins and Self Joins in Limited Storage', ACM PODS Conference, 1999.

\bibitem{anglano}
Anglano, C., Canonico, M., Guazzone, M., 'Profit-aware Resource Management for Edge Computing Systems', 1st International Workshop on Edge Systems, Analytics and Networking, 2018, pp. 25--30.

\bibitem{Babcock1}
Babcock, B., Datar, M., Motwani, R., 'Load Shedding Techniques for Data Stream Systems', Workshop on Management and Processing of Data Streams, 2003.

\bibitem{Babcock}
Babcock, B., Babu, S., Datar, M., Motwani, R., Widom, J., 'Models and issues in data stream systems', PODS, 2002.

\bibitem{bellavista}
Bellavista, P., Corradi, A., Foschini, L., Scotece, D., 'Differentiated Service/Data Migration for Edge Services Leveraging Container Characteristics', IEEE Access, vol. 7, 2019.

\bibitem{ref32}
Bhardwaj, K., Agrawal, P., Gavrilovska, A., Schwan, K., 'AppSachet: Distributed App Delivery from the Edge Cloud', 7th International Conference Mobile Computing, Applications, and Services, 2015, pp. 89-–106.

\bibitem{Chakrabarti}
Chakrabarti K., Garofalakis M., Rastogi R., Shim, K., 'Approximate Query Processing with Wavelets', VLDB Journal, vol. 10(2-3), 2001, pp. 199--223.

\bibitem{Charikar}
Charikar M., Chen K., Farach-Colton, M., 'Finding Frequent items in data streams', ICALP, 2002.

\bibitem{Cherrueau}
Cherrueau, R. A., Lebre, A., Pertin, D., Wuhib, F., Soares, J., 'Edge Computing Resource Management System: a Critical Building Block! Initiating the debate via OpenStack', USENIX Workshop on Hot Topics in Edge Computing, 2018, pp. 1--6.

\bibitem{chu}
Chu, D., Deshpande, A., Hellerstein, J., Hong, W., 'Approximate Data Collection in Sensor Networks using Probabilistic Models', in 22nd International Conference on Data Engineering (ICDE'06), 2006.

\bibitem{Cormode}
Cormode G., Muthukrishnan, S., 'What's hot and what's not: Tracking most frequent items dynamically', ACM PODS Conference, 2003.

\bibitem{dobra}
Dobra A., Garofalakis M. N., Gehrke J., Rastogi, R., 'Sketch-Based Multi-query Processing over Data Streams', EDBT Conference, 2004.

\bibitem{Gehrke}
Gehrke J., Korn, F., Srivastava, D., 'On Computing Correlated Aggregates Over Continual Data Streams', SIGMOD Conference, 2001.

\bibitem{ref28}
Hagras, H., 'A hierarchical type-2 fuzzy logic control architecture for autonomous mobile robots', IEEE TFS, vol. 12, 2004.

\bibitem{holt}
Holt, C., 'Forecasting seasonals and trends by exponentially weighted moving averages', International Journal of Forecasting, vol 20(1), 2004, pp. 5--10.

\bibitem{kolomvatsos2017}
Kolomvatsos, K., Anagnostopoulos, C., Marnerides, A., Ni, Q., Hadjiefthymiades, S., Pezaros, D., 'Uncertainty-driven Ensemble Forecasting of QoS in Software Defined Networks', 22nd IEEE Symposium on Computers and Communications (ISCC), Heraklion, Greece, 2017.

\bibitem{kolomvatsos2016}
Kolomvatsos, K., Anagnostopoulos, C., Hadjiefthymiades, S., 'Data Fusion \& Type-2 Fuzzy Inference in Contextual Data Stream Monitoring', IEEE Transactions on Systems, Man and Cybernetics: Systems, vol. PP, Issue 99, pp.1-15, 2016.

\bibitem{Lakshmi}
Lakshmi, K. P., Reddy, C. R. K., 'A Survey on Different Trends in Data Streams', IEEE International Conference on Networking and Information Technology, 2010.

\bibitem{Manku}
Manku, G., Motwani, R., 'Approximate Frequency Counts over Data Streams', VLDB Conference, 2002.

\bibitem{martin}
Martin, R., Vahdat, A., Culler, D., Anderson, T., 'Effects of Communication Latency, Overhead, and Bandwidth in a Cluster Architecture', 4th Annual International Symposium on Computer Architecture, 1997.

\bibitem{ref44}
Mendel, J. M., 'Type-2 Fuzzy Sets and Systems: An Overview', IEEE Computational Intelligence Magazine, 2(2), 2007.

\bibitem{ref45}
Mendel, J. M., 'Uncertain Rule-Based Fuzzy Logic Systems: Introduction and New Directions', Upper Saddle River, Prentice-Hall, 2001.

\bibitem{mesiar} 
Mesiar, R., Kolesarova, A., Calvo, T., Komornikova, M., 'A Review of Aggregation Functions',
Studies in Fuzziness and Soft Computing, 2008.

\bibitem{Muthukrishnan}
Muthukrishnan, S., 'Data streams: algorithms and  applications', 
14th annual ACM-SIAM symposium on discrete algorithms, 2003.

\bibitem{najam}
Najam, S., Gilani, S., Ahmed, E., Yaqoob, I., Imran, M.,  'The Role of Edge Computing in Internet of Things', IEEE Communications Magazine, 2018, doi: 10.1109/MCOM.2018.1700906.

\bibitem{fuzzy}
Novák, V., Perfilieva, I., Močkoř, J., 'Mathematical principles of fuzzy logic', Dordrecht: Kluwer Academic, 1999.

\bibitem{ref26}
Sardellitti, S., Scutari, G., Barbarossa, S., 'Joint  Optimisation of Radio and Computational Resources for Multicell Mobile-Edge Computing',
IEEE Transactions on Signal and Information Processing over Networks, vol. 1(2), 2015, pp. 89-–103.

\bibitem{ref33}
Savolainen, P., Helal, S., Reitmaa, J., Kuikkaniemi, K., Jacucci, G., Rinne, M.,   Turpeinen, M., Tarkoma, S., 'Spaceify:  A  Client-edge-server  Ecosystem  for  Mobile  Computing  in  Smart  Spaces', International Conference on Mobile Computing \& Networking, 2013, pp. 211–-214.

\bibitem{Schweller}
Schweller, R., Gupta, A., Parsons, E., Chen, Y., 'Reversible Sketches for Efficient and Accurate Change Detection over Network Data Streams', Internet Measurement Conference Proceedings, 2004.

\bibitem{Shekhar}
Shekhar, S., Gokhale, A., 'Dynamic Resource Management Across Cloud-Edge Resources for Performance-Sensitive Applications', 17th IEEE/ACM International Symposium on Cluster, Cloud and Grid Computing, 2017.

\bibitem{ref34}
Simoens, P., Xiao, Y., Pillai, P., Chen, Z., Ha, K., Satyanarayanan, 'Scalable crowd-sourcing of video from mobile devices', 11th annual international conference on Mobile systems, applications, and services, 2013, pp. 139--152.

\bibitem{Tatbul}
Tatbul, N., Zdonik, S., 'A subset-based load shedding approach for aggregation queries over data streams', International Conference on Very Large Data Bases (VLDB), 2006.

\bibitem{wang}
Wang, N., Varghese, B., Matthaiou, M., Nikolopoulos, D., 'ENORM: A Framework for Edge Node Resource Management', IEEE Transactions on Service Computing, 2017, doi: 10.1109/TSC.2017.2753775.

\bibitem{ref17}
Yao, Y., Cao, Q., Vasilakos, A. V., 'EDAL: An Energy-Efficient,Delay-Aware, and Lifetime-Balancing Data Collection Protocol for Wireless Sensor Networks', IEEE International
Conference on Mobile Ad-Hoc and Sensor Systems, 2013, pp. 182-–190.

\bibitem{ref21}
Zhou, A., Wang, S., Li, J., Sun, Q., Yang, F., 'Optimal  Mobile Device Selection for Mobile Cloud Service Providing', The Journal of Supercomputing, vol. 72(8), 2016, pp. 3222-–3235.

\end{thebibliography}
\end{document}